\documentclass[oribibl]{llncs}

\usepackage[colorlinks=true,
  linkcolor=blue,citecolor=blue,urlcolor=blue]{hyperref}

\usepackage{url}
\usepackage{graphicx}

\newcommand{\Qa}{$\textmd{Q}_1$}
\newcommand{\Qb}{$\textmd{Q}_2$}

\usepackage{fancyhdr} 

\fancypagestyle{firststyle}
{
   \fancyhf{}
   \fancyfoot[C]{\scriptsize{Proceedings of the First International Workshop ``CAiSE for Legal Documents'' (COUrT 2020), Grenoble (online), CEUR-WS, pp.~3--14. The final publication is available via \url{http://ceur-ws.org/Vol-2690/COUrT-paper1.pdf}}}
}

\begin{document}

\title{Predicting the Amount of GDPR Fines}

\author{Jukka Ruohonen \and Kalle Hjerppe
\\\email{\{juanruo, kphjer\}@utu.fi}}
\institute{Department of Future Technologies, University of Turku, Turku, Finland}

\maketitle

\begin{abstract}
The General Data Protection Regulation (GDPR) was enforced in 2018. After this
enforcement, many fines have already been imposed by national data protection
authorities in the European Union (EU). This paper examines the individual GDPR
articles referenced in the enforcement decisions, as well as predicts the amount
of enforcement fines with available meta-data and text mining features extracted
from the enforcement decision documents. According to the results, articles
related to the general principles, lawfulness, and information security have
been the most frequently referenced ones. Although the amount of fines imposed
vary across the articles referenced, these three particular articles do not
stand out. Furthermore, good predictions are attainable even with simple machine
learning techniques for regression analysis. Basic meta-data (such as the
articles referenced and the country of origin) yields slightly better
performance compared to the text mining features.
\end{abstract}


\section{Introduction}

\thispagestyle{firststyle} 

Data protection has a long history in the EU. In particular, the GDPR repealed
the earlier Directive 95/46/EC. Although this directive laid down much of the
legal groundwork for EU-wide data protection and privacy, its national
adaptations, legal interpretations, and enforcement varied both across the
member states and different EU institutions~\cite{Fuster14}. In short: it was a
paper tiger. In contrast, Regulation (EU) 2016/679, the GDPR, is a regulation;
it is binding throughout the EU with only a minimal space for national
adaptations. In practice, only a few Articles~(A) in the GDPR provide some but
limited room for national maneuvering; these include A6 with respect to
relaxation in terms of other legal obligations or public interests, A9 in terms
of sensitive data, and A10 regarding criminal matters. Thus, in general, this
particular legislation should be interpreted and enforced uniformly through the
European Union by national data protection authorities whose \textit{formal}
powers are defined in A58. In practice, however, already the resources and thus
the \textit{actual} power for enforcement vary across the member
states~\cite{Bennett18a, Custers18}. Coupled with a lack of previous research on
the enforcement of the GDPR, this variance provides a motivation for the present
work to examine the recent enforcement fines imposed according to the conditions
specified in A83. In addition, the work is motivated by a tangential question;
is it also possible to predict these fines by machine~learning~methods?

To answer to the question, the paper uses meta-data and text miming features
extracted from the decision documents released by the national authorities. As
such, only black-box predictions are sought; the goal is not to make any legal
interpretations whatsoever. Nevertheless, the answer provided still establishes
a solid contribution---especially when considering that the paper is presumably
the very first to even examine the GDPR fines. As is discussed in
Section~\ref{sec: background}, the black-box approach also places the paper into
a specific branch of existing research dealing with legal documents. This
section also refines the question into two more specific research
questions. Afterwards, the structure is straightforward: the dataset and methods
are elaborated in Sections~\ref{sec: data} and \ref{sec: methods}, results are
presented in Section~\ref{sec: results}, and conclusions follow in
Section~\ref{sec: conclusion}. As will be noted in the final section, there are
also some lessons that should \textit{not} be learned from this~work.

\section{Background}\label{sec: background}

Legal mining---in lack of a better term---has emerged in recent years as a
promising but at times highly contested interdisciplinary field that uses
machine learning techniques to analyze various aspects related to
law~\cite{Dyevre19}. Although the concrete application domains vary, case law
and court cases are the prime examples already because these constitute the
traditional kernel of legal scholarship. Within this kernel, existing machine
learning applications range from the classification of judges' ideological
positions \cite{Hausladen20}, which may be illegal in some European
countries~\cite{CITIP20a}, to the prediction of decisions of the European Court
of Human Rights~\cite{LiuChen17, Medvedeva19}. These examples convey the
traditional functions of applied machine learning; exploratory data mining and
the prediction of the future.

There is also another closely related application domain. Again in lack of a
better term, data extraction could be a label for this domain: by exploiting the
nature of law as an art of persuasion~\cite{Dyevre19}, the domain uses distinct
information retrieval techniques to extract and quantify textual data from legal
documents into structured collections with a predefined logic and
semantics~\cite{Breaux06, Sleimi19, Wagh20}. To gain a hint about the
extraction, one might consider a legal document to contain some facts, rights,
obligations, and prohibitions, statements and modalities about these, and so
forth. Although the two application domains are complementary in many respects,
the underlying rationales exhibit some notable differences.

Oftentimes, the legal mining domain is motivated by a traditional rationale for
empirical social science research: to better understand trends and patterns in
lawmaking and law enforcement; to contrast these with legal philosophies and
theories; and so forth. This rationale extends to public administration: machine
learning may ease the systematic archiving of legal documents and the finding of
relevant documents, and, therefore, it may also reduce administrative
costs~\cite{Chatwal17}. These administrative aspects reflect the goal of
building ``systems that assist in decision-making'', whereas the predictive
legal mining applications seek to build ``systems that make
decision''~\cite{Nissan18}. Although the data extraction domain can be motivated
by the same administrative rationale, providing data to predictive systems is
seldom the intention behind the extraction. Instead, there is a further
rationale in this domain: to extract requirements for software and systems in
order to comply with the laws from which a given extraction is
done~\cite{Sleimi19}. Driven by the genuine interest to facilitate collaboration
between lawyers and engineers in order to build law-compliant software and
systems~\cite{vanDijk18}, this rationale has been particularly prevalent in the
contexts of data protection and privacy. For instance, previous work has been
done to extract requirements from the Health Insurance Portability and
Accountability Act in the United States~\cite{Breaux06}. Against this backdrop,
it is no real surprise that data extraction has been applied also for laws
enacted in the EU. While there is previous work for identifying requirements
from the GDPR manually~\cite{Hjerppe19a}, there indeed exists also more
systematic data extraction approaches~\cite{Tamburri20}. However, neither domain
has addressed the enforcement of this EU-wide regulation. In fact, a reasonably
comprehensive literature search indicates no previous empirical research on the
GDPR's enforcement. Given this pronounced gap in the existing literature, this
paper sets to examine the following two Questions (Q) regarding the enforcement
fines:
\begin{description}
\itemsep 2pt
\item{\Qa:~\textit{(i)~Which GDPR articles have been most often referenced in the recent enforcement cases, (ii) and do the enforcement fines vary across these articles?}}
\item{\Qb:~\textit{How well the recent GDPR fines can be predicted in terms of basic available (i)~meta-data and (ii)~textual traits derived from the enforcement decisions?}}
\end{description}

These two questions place the present work into the legal mining domain. Also the underlying rationales are transferable. For instance, an answer to \Qa~helps to understand which aspects of the GDPR have been actively enforced during the early roll out of the regulation. Also \Qb~carries a practical motivation: by knowing whether the penalties are predictable by machine learning techniques, a starting point is available for providing further insights in different practical scenarios. These scenarios range from the automated archival of enforcement decisions and the designation of preventive measures to litigation preparations. However, it is important to remark that the GDPR's enforcement is done by national data protection authorities. Although the focus on public administration is maintained nevertheless, documents about the enforcement decisions reached by these authorities should not be strictly equated to law-like legal documents. This point provides an impetus to move forward by elaborating the dataset used.

\section{Data}\label{sec: data}

The dataset is based on a GDPR enforcement tracker that archives the fines and
penalties imposed by the European data protection
authorities~\cite{enforcementtracker20}. This tracker is maintained by an
international law firm for archiving many of the known enforcement cases. Each
case is accompanied by meta-data supplied by the firm as well as a link to the
corresponding decision from a national authority. In addition to potentially
missing cases due to the lack of publicly available information, the archival
material is unfortunately incomplete in many respects. The reason originates
from the incoherent reporting practices of the European data protection
authorities. Therefore, all cases were obtained from the tracker, but the
following four steps were followed to construct a sample for the analysis:
\begin{enumerate}
\itemsep 2pt
\item{To maintain coherence between \Qa~and~\Qb, only those cases were included
  that had both meta-data and links to the decisions available. In terms of the
  former, some cases lacked meta-data about the fines imposed, the particular
  GDPR articles referenced in the decisions, and even links to the decisions.}
\item{To increase the quality of the sample, only those cases were included that
  were accompanied with more or less formal documents supplied on the official
  websites of the data protection authorities. Thus, those cases are excluded
  whose archival material is based online media articles, excerpts collected
  from annual reports released by the authorities, and related informal
  sources.}
\item{If two or more cases were referenced with the same decision, only one
  decision document was included but the associated meta-data was unified into a
  single case by merging the articles references and totaling the fines
  imposed.}
\item{All national decisions written in languages other than English were
  translated to English with Google Translate. In general, such machine
  translation is necessary due to the EU-wide focus of the forthcoming empirical
  analysis.}
\end{enumerate}

Given these restrictions, the sample amounts to about 72\% of all cases archived
to the tracker at the time of data collection. Even with these precautions, it
should be stressed that the quality of the sample is hardly optimal. While the
accuracy of the meta-data supplied by the firm is taken for granted, there are
also some issues with the quality of the publicly available decisions. The
authorities in some countries (e.g., Hungary and Spain) have released highly
detailed and rigorous documents about their decisions, while some other
authorities (e.g., in Germany) have opted for short press releases. Although
most of the documents were supplied in the portable document format (PDF) and
informally signed by the authorities, it should be thus stressed that the data
quality is not consistent across the European countries observed. In addition,
it is worth remarking the detail that scanned PDF documents (as used, e.g., in
Portugal) had to be excluded due to the automatic data processing. While these
data quality issues underline the paper's exploratory approach, these carry also
political and administrative ramifications that are briefly discussed later on
in Section~\ref{sec: conclusion}.

\section{Methods}\label{sec: methods}

Descriptive statistics and regression analysis are used for answering to the two
questions asked. In terms of Question~\Qa, dummy variables for the GDPR articles
referenced are simply regressed against the logarithm of the fines imposed by
using the conventional analysis-of-variance (ANOVA). As many of the cases
reference multiple articles, it should be remarked that these dummy variables
are not so-called fixed effects. The methods for answering to the second
Question~\Qb~require a more thorough elaboration. In addition to (i)~the GDPR
\textit{articles}, the meta-data aspects include dummy variables for the
following features: (ii)~the \textit{year} of a given enforcement case; (iii)
the \textit{country} in which the given fine was imposed; and (iv) the
\textit{sector} of the violating organization. The last feature was constructed
manually by using five categories: individuals, public sector (including
associations), telecommunications, private sector (excluding
telecommunications), and unknown sector due to the lack of meta-data supplied in
the enforcement tracker. In total, these features amount to $49$ dummy
variables.

The textual aspects for \Qb~are derived from the translated decisions. Seven
steps were used for pre-processing: (a) all translated decision documents were
lower-cased and (b) tokenized according to white space and punctuation
characters; (c) only alphabetical tokens recognized as English words were
included; (d)~common and custom stopwords were excluded; (e) tokens with lengths
less than three characters or more than twenty characters were excluded; (f) all
tokens were lemmatized into their common English dictionary forms; and, finally,
(g) those lemmatized tokens were excluded that occurred in the whole decision
corpus in less than three times. A common natural language processing
library~\cite{NLTK19} was used for this processing together with a common
English dictionary~\cite{hunspell20}. In addition to the stopwords supplied in
the library, the twelve most frequent tokens were used as custom excluded
stopwords: \textit{data}, \textit{article}, \textit{personal},
\textit{protection}, \textit{processing}, \textit{company}, \textit{authority},
\textit{regulation}, \textit{information}, \textit{case}, \textit{art}, and
\textit{page}. After this pre-processing, the token-based term frequency (TF)
and term frequency inverse document frequency (TF-IDF) were calculated from the
whole corpus constructed (for the exact formulas used see, e.g.,
\cite{Ruohonen18TIR}). These common information retrieval statistics are used
for evaluating the other part in~\Qb. In general, TF-IDF is often preferred as
it penalizes frequently occurring terms.

Sparsity is the biggest issue for prediction. There are only $154$ observations
but already the meta-data amounts to $49$ independent variables---and the TF and
TF-IDF each to $4189$ independent variables. Fortunately, the problem is not
uncommon, and well-known solutions exist for addressing it. Genomics is a good
example about the application domains riddled with the problem; within this
domain, it is not uncommon to operate with datasets containing a few thousand
observations and tens of thousands of predictors~\cite{Colombani12}. Dimension
reduction is the generic solution in this domain and other domains with similar
problems. Thus, three common dimension reduction methods for regression analysis
are used: principal component regression (PCR), partial least squares (PLS), and
ridge regression (for a concise overview of these methods see, e.g.,
\cite{Hastie11}). In essence, PCR uses uncorrelated linear combinations as the
independent variables; PLS is otherwise similar but also the dependent variable
is used for constructing the combinations. Ridge regression is based on a
different principle: the dimensionality is reduced by shrinking some of the
regression coefficients to zero. In general, all three methods are known to
yield relatively similar results in applied work.

In terms of practical computation, the number of components for the PCR and PLS
models, and the shrinkage parameter for the ridge regression, is optimized
during the training while the results are reported with respect to a test set
containing 20\% of the enforcement cases. Centering (but not scaling) is used
prior to the training with a $5$-fold cross-validation. Computation is carried
out with the \textit{caret} package~\cite{caret20} in conjunction with the
\textit{pls}~\cite{pls07} and \textit{foba} \cite{foba08} packages. Although
root-mean-square errors (RMSEs) are used for optimizing the training, the
results are summarized with mean absolute errors (MAEs) due to their
straightforward interpretability. These are defined as the arithmetic means of
the absolute differences between the observed and predicted fines in the
test~set.

\section{Results}\label{sec: results}

\begin{figure}[th!b]
\centering
\includegraphics[width=11cm, height=3.5cm]{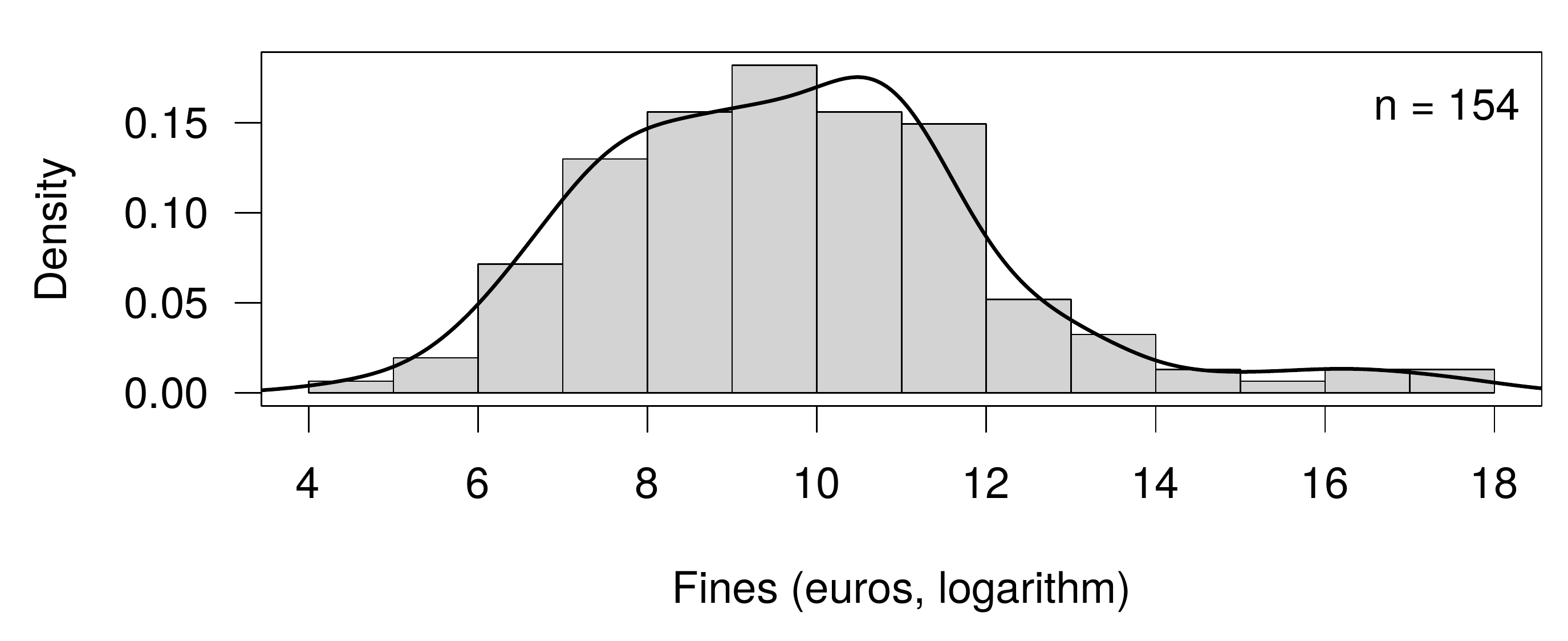}
\caption{Enforcement Fines in the Sample}
\label{fig: fines}
\end{figure}

The GDPR fines imposed vary greatly. As can be seen from Fig.~\ref{fig: fines},
a range from about $e^6$ euros to $e^{12}$ euros capture the majority of the
enforcement fines observed. This range amounts roughly from about four hundred
to $163$ thousand euros. That said, the distribution has a fairly long tail;
also a few large, multi-million euro fines are present in the sample. Therefore,
the sample cannot be considered biased even though the restrictions discussed in
Section~\ref{sec: data} exclude some of the largest enforcement cases, including
the announcements about the intention to fine the British Airways and Marriott
International by the Information Commissioner's Office in the United
Kingdom. Although these two excluded cases are---at least at the time of
writing---preliminary announcements, they are still illuminating in the sense
that both were about large-scale data breaches.

\begin{figure}[th!b]
\centering
\includegraphics[width=\linewidth, height=3.5cm]{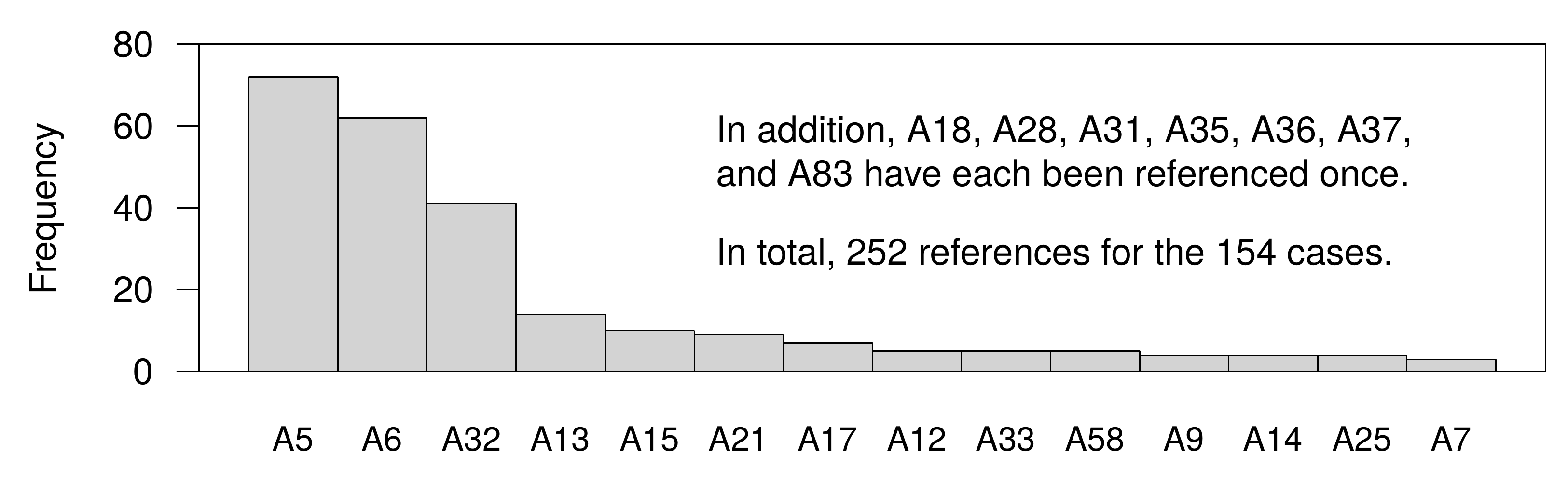}
\caption{Referenced GDPR Articles in the Enforcement Cases}
\label{fig: articles}
\end{figure}

However, the GDPR's corresponding A32 for information security has not been the
most frequently referenced article in the recent enforcement cases. Instead, A5
and A6, which address the general principles and lawfulness of personal data
processing, have clearly been the most referenced individual articles, as can be
seen from Fig.~\ref{fig: articles}. These two articles account for as much as
87\% of all $252$ references made in the $154$ enforcement cases. More than six
references have been made to A13 (informing obligations to data subjects), A15
(right to access), A21 (right to object), and A17 (right to erasure). These
references indicate that enforcement has been active also with respect to the
rights granted by the GDPR for individual data subjects. Furthermore, less
frequent references have been made in the decisions to numerous other
articles. These include the obligations to designate data protection officers
(A37), conduct impact assessments (A35), and consult supervisory authorities
(A36), to name three examples. While the principles, lawfulness, and information
security account for the majority, the less frequent but still visible
references to more specific articles hint that the regulation's whole scope is
slowly being enforced by the European authorities.

\begin{figure}[t!]
\centering
\includegraphics[width=\linewidth, height=4cm]{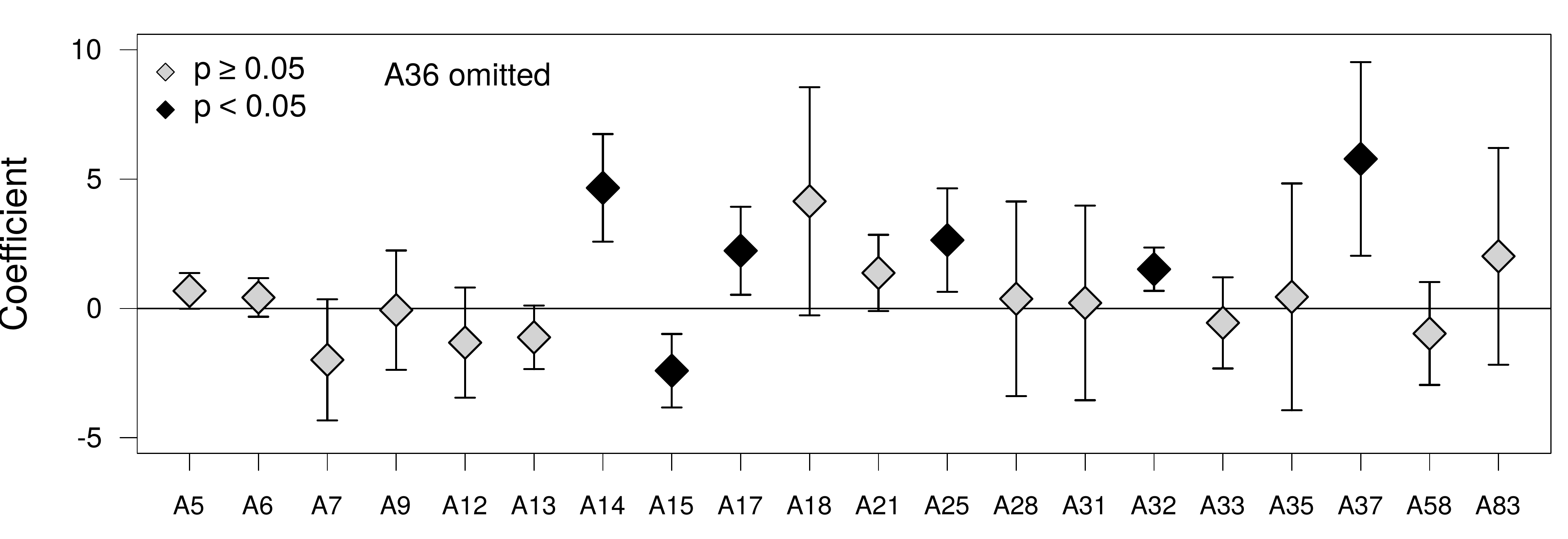}
\caption{Enforcement Fines Across Articles (logarithm, ANOVA, 95\% CIs)}
\label{fig: articles fines}
\end{figure}

Turning to the second part of \Qa, the regression coefficients from the
log-linear ANOVA model are visualized in Fig.~\ref{fig: articles fines} (the
intercept is present in the model but not shown in the figure, and A36 is
omitted as the single reference made to the article corresponds with the single
reference made to A35 in the same decision; the dummy variable for A35 thus
captures the effect of both articles). As can be seen, the confidence intervals
(CIs) are quite wide for the articles referenced only infrequently, and only six
coefficients are statistically significant at the conventional threshold. Thus,
some care is required for interpretation.

When looking at the coefficients with relatively tight CIs, it is evident that
variation is present but the magnitude of this variation is not
substantial. Most of the coefficients remain in the range $[-5, 5]$. However,
together all the references do yield a decent model; an $F$-test is
statistically significant and the coefficient of determination is large ($R^2
\simeq 0.44$). To put aside the statistical insignificance, it is also
interesting to observe that some of the coefficients have negative signs,
meaning that some references indicate smaller fines compared to the
average. Among these are the conditions for consent (A7), sensitive data (A9),
transparency (A12), and informing (A13), as well as the already noted right to
access (A15), proper notifications about data breaches (A33), and the powers
granted for the supervisory authorities (A58). Finally, the magnitude of the
coefficient ($1.52$) for the information security article (A32) is significant
but does not stand out in terms of magnitude. When compared to cases without a
reference to this article, only about $1.5\%$ higher fines have been imposed in
cases referencing A32.

\begin{figure}[t!]
\centering
\includegraphics[width=\linewidth, height=9cm]{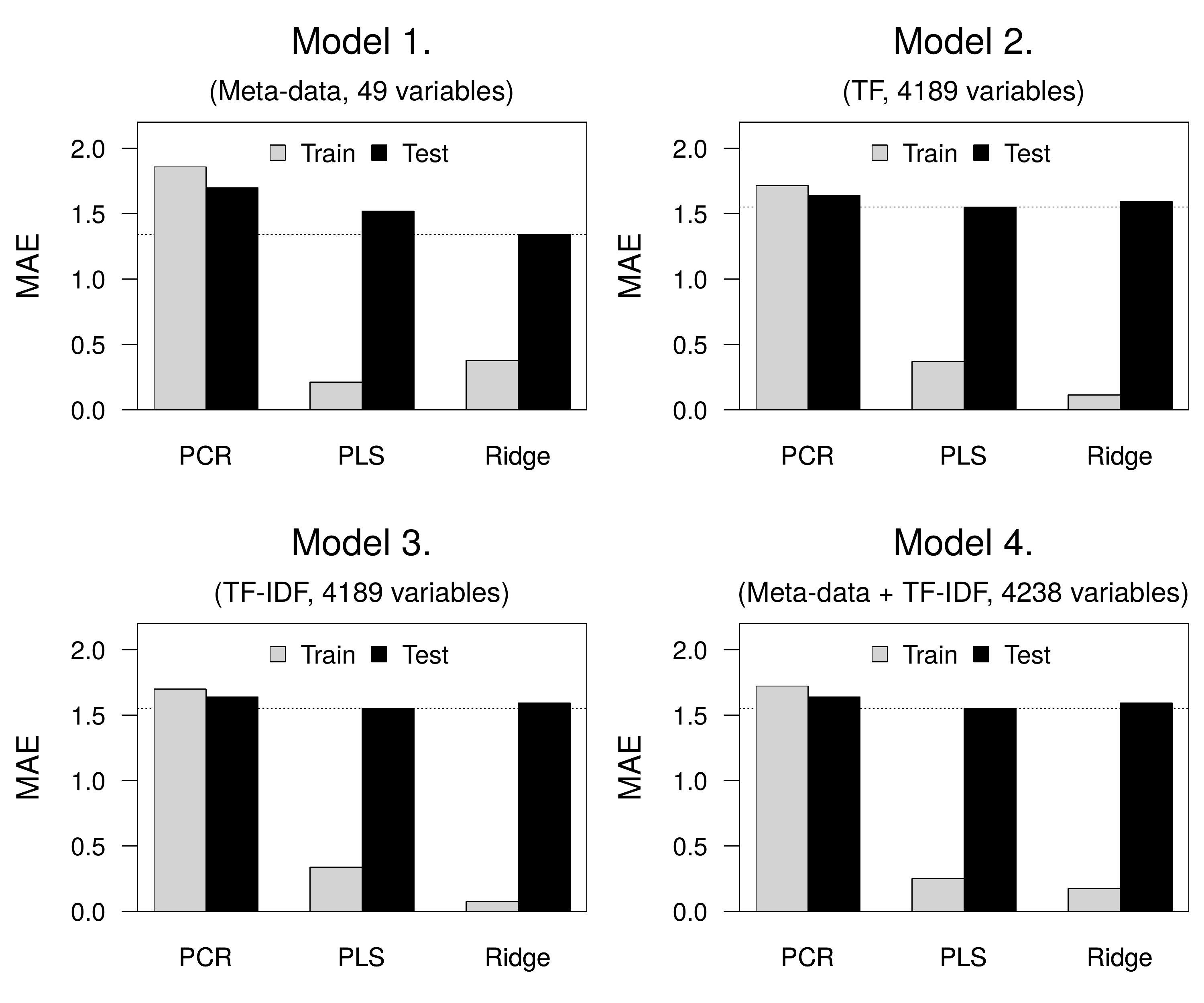}
\caption{Prediction Performance (logarithm, MAEs)}
\label{fig: performance}
\end{figure}

\begin{figure}[t!]
\centering
\includegraphics[width=\linewidth, height=4cm]{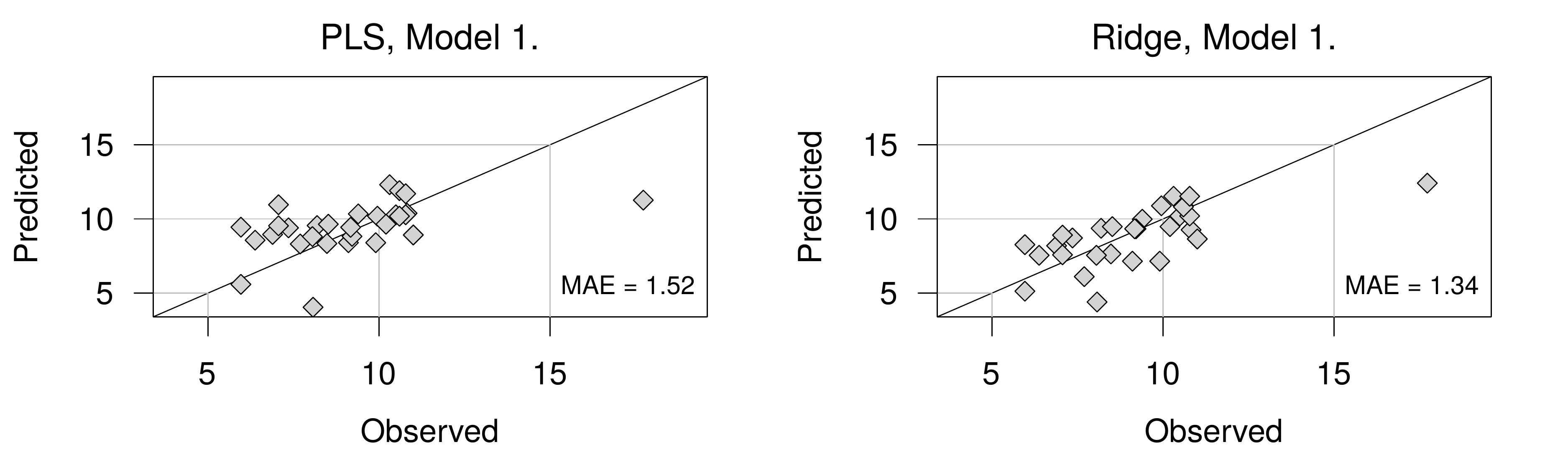}
\caption{Observed and Predicted Values in the Test Set}
\label{fig: predictions}
\end{figure}

The results regarding \Qb~are summarized in Fig.~\ref{fig: performance} (the
MAEs for the training refer to the best cross-validated models). Three
noteworthy observations can be drawn from this summary. First and foremost, the
prediction performance is generally decent: the best-performing cases all yield
MAEs roughly between $1.3$ and $1.5$ for the log-transformed fines. These
average prediction errors seem also reasonable when taking a closer look at the
actual predictions---except for the outlying large fines. Take Fig.~\ref{fig:
  predictions} as a brief example; the figure displays the observed fines and
the predicted fines based on the PLS and ridge regression estimators for the
first meta-data model. Even though most of the predicted observations are fairly
close to the observed fines, the test set also contains one five million euro
fine that is quite severely underestimated by both regression estimators. The
underestimations amount to over $246$ thousand euros. Though, when a magnitude
is measured in millions, it is a matter of interpretation whether an error
measured in hundreds of thousands is large, small, or something else.

Second, there are some interesting differences between the regression
estimators. In particular, PLS and ridge regression exhibit relatively large
differences between training and testing. The explanation relates to the
RMSE-based optimization during training. For instance, PCR was estimated with
only one component for the first meta-data model and three components for the
remaining three models, whereas two components were picked for all four
PLS~models.

Last but not least, the smallest MAE for the test set is outputted by ridge
regression using only the $49$ meta-data variables. The second and third models
containing the TF and TF-IDF variables both perform worse. Furthermore, the
fourth model, which contains the meta-data and TF-IDF variables, indicates that
the text mining features tend to slightly weaken the predictions. It is also
worth remarking that some redundancy is present among the meta-data variables;
comparable performance is obtained with only $17$ meta-data variables that are
left after prior pre-processing with the \textit{caret}'s \texttt{nearZeroVar}
function. All this said, the overall interpretation should be less explicit when
considering the practical motivation for \Qb~noted in Section~\ref{sec:
  background}. If only the decision documents are available without any prior
work to manually construct the meta-data from these, even the simple text mining
features could be used for black-box predictions.

\section{Conclusion}\label{sec: conclusion}

This paper explored two questions. The answers to these can be summarized as
follows. First: regarding \Qa, the articles related to the general principles
(A5), lawfulness (A6), and information security (A32) have been most frequently
referenced by the national data protection authorities during the early
enforcement period observed in this paper. Although also the enforcement fines
vary across the various GDPR articles referenced in the authorities' decisions,
the effects of these three articles do not stand out in particular. A good
corollary question for further work would be to examine the future evolution of
these references; a hypothesis is that the regulation's enforcement is slowly
moving from the principles and lawfulness conditions to more specific
elements. Then: regarding~\Qb, it is possible to obtain decent predictions even
with standard machine learning techniques for regression analysis. Basic
meta-data (i.e., articles referenced, year of enforcement, country or origin,
and industry sector) seems to provide slightly better predictive performance
compared to basic text mining features (i.e., TF and TF-IDF) extracted from the
decision documents. Yet, even the text mining features seem sufficient for blind
black-box predictions. There are also many potential ways to improve the
predictions reported, including those related regression analysis (such as using
specific sparse-PLS estimators) and text mining (such as using word
embeddings). Data mining techniques (such as topic modeling) could be used also
for better understanding the nuances behind the decisions. An alternative path
forward would be to extend the specific data extraction approaches discussed in
Section~\ref{sec: background} to the enforcement decisions. However, the
motivation to move forward is undermined by practical problems. As was remarked
in Section~\ref{sec: data}, already the quality of data is a problem of its own.

Recently, the enforcement of the GDPR has been fiercely criticized by some
public authorities and pundits alike. The reasons are many: a lack of
transparency and cooperation between national data protection authorities,
diverging legal interpretations, cultural conflicts, the so-called
``one-stop-shop'' system, old-fashioned information systems and poor data
exchange practices, and so on and so forth~\cite{Politico19a}. The data
collection used for the present work testifies on behalf of the criticism: the
decision documents released by the national authorities have varied wildly in
terms of quality and rigor. Some national authorities have even hidden their
decisions from public scrutiny.  A paradox is present: although A15 grants a
right for data subjects to access their personal data, the same subjects may
need to exercise their separate freedom of information rights to obtain cues
about decisions reached by national authorities. Four legs good, two legs bad.

Finally, it is necessary to briefly point out the bigger issues affecting the
legal mining and data extraction domains---and, therefore, also the present
work. For one thing, the practical usefulness of legal expert systems has been
questioned for a long time. The artificial intelligence hype has not silenced
the criticism~\cite{Leith16}. Like with the ``code is law'' notion, which has
never existed in reality~\cite{MuellerBadiei19}, there are also many
philosophical counterarguments against the legal mining and data extraction
domains~\cite{Dyevre19, Franklin12, Nissan18}. It is problematic at best to
codify the methodology of a scholarly discipline into rigid schemas in order to
nurse the methodological requirements of another discipline; legal reasoning is
distinct from other types of reasoning exercised in empirical sciences; and so
forth. Law is not code. But code is increasingly used to predict law enforcement
decisions. The legal mining domain, in particular, is frequently involved with a
motivation to build ``a~system that could predict judicial decisions
automatically'' but with a provision that there is ``no intention of creating a
system that could replace judges'' \cite{Medvedeva19}. Such system-building
leads to another delicate paradox. Namely, the GDPR and related laws (such as
Directive 2016/680 for data protection in criminal matters) were also designed
to provide certain guards \textit{against} legal mining and the resulting
automated decision-making involving human beings~\cite{Zavrsnik20}. This paper
is not immune to criticism originating from this fundamental paradox. If it is
seen as undesirable to build systems for making law enforcement decisions, it
should be also seen as undesirable to build systems for automatically
fining~companies.

\subsection*{Acknowledgements}

\enlargethispage{0.4cm}
This research was funded by the Strategic Research Council at the Academy of Finland (grant no.~327391).

\bibliographystyle{splncs03}

\end{document}